\documentclass[aps,prb,reprint,twocolumn,showpacs,floatfix,superscriptaddress]{revtex4}
\usepackage{graphicx}
\usepackage{amssymb}
\usepackage{amsmath}
\usepackage{lmodern}
\usepackage{color}
\usepackage{hyperref}
\usepackage{simplewick}
\begin{document}

\preprint{This line only printed with preprint option}

\title{Critical behaviour of anisotropic  magnets with quenched disorder: replica symmetry breaking  studied by  operator product expansion}

\author{E. Kogan}

\affiliation{Jack and Pearl Resnick Institute, Department of Physics, Bar-Ilan University, Ramat-Gan 52900, Israel}
\affiliation{Max-Planck-Institut fur Physik komplexer Systeme,  Dresden 01187, Germany}
\affiliation{Center for Theoretical Physics of Complex Systems, Institute for Basic Science (IBS), Daejeon 34051, Republic of Korea}

\author{M. Kaveh}

\affiliation{Jack and Pearl Resnick Institute, Department of Physics, Bar-Ilan University, Ramat-Gan 52900, Israel}
\affiliation{Cavendish Laboratory, University of Cambridge, J J Thomson Avenue, Cambridge CB3 0HE, UK}

\date{\today}

\begin{abstract}
We study critical behaviour of  disordered magnets near four dimensions.
We consider  the system with explicit cubic anisotropy and
scalar disorder and that  with random direction of anisotropy axis. The quenched disorder is taken into account by replica method.
Using the
method of operator product expansion, we derive in the first order to $\epsilon$ approximation the renormalization group  equations
taking into account possible replica symmetry breaking.
\end{abstract}

\pacs{64.60.ae,64.60.Ej,64.60.F-}

\maketitle

\section{Introduction}

Wilson and Fisher calculation of the critical exponents for the $\phi^4$ model by $\epsilon$ expansion of the renormalization group (RG) equations,\cite{wilson} was a groundbreaking discovery.
Soon after that the $\epsilon$ expansion   was used
 to derive the RG equations for the $\phi^4$-model with  scalar quenched disorder \cite{khmel,lubensky,grinstein}  (see also  Ref. \onlinecite{ma}).
Anisotropic models were studied by Aharony:  first the pure $\phi^4$ model  \cite{aharony}  (see also Refs. \onlinecite{chaikin,izyumov,cardy}), and later disordered anisotropic models \cite{aharony2}.

The quenched disorder was taken into account in the mentioned above works by the replica  method \cite{dotsenko}. This method was advanced by Dotsenko et al. \cite{dotsenko2}, who
have shown that the replica symmetry, assumed in the previous application of the method to the RG theory, can be spontaneously broken.
(This kind of replica symmetry breaking (RSB) was previously discovered by Parisi in the theory of spin glasses \cite{parisi}.)

The two  models one with  cubic anisotropy and quenched scalar disorder
and another with random direction of the axis of anisotropy, studied previously by Aharony, is the subject of the present contribution. The RG equations are obtained in a very simple
and appealing way using the operator product expansion (OPE) method, another great discovery of Wilson \cite{wilson2} (see also Refs. \onlinecite{peskin,holland}; application of this method
 to the theory of classical phase transitions is particularly clearly presented  in the  book by Cardy \cite{cardy}).
To present the method and notation we start by rederiving by the OPE method the    replica symmetric RG equations. In the most important in the paper Section \ref{rsb} we generalize
these equations to take into account the  RSB.

\section{Operator product expansion and the perturbative renormalization group}
\label{pert}

The operator product expansion is a  universal conception of quantum field theory.
The essential idea is that for any two local operator quantum fields at  points ${\bf x},{\bf y}$ (we consider Euclidean space) their product may be
expressed in terms of a series of local quantum fields at any other point  ( which may be identified with ${\bf x}$ or ${\bf y}$)
times $c$-number coefficient functions which depend on $|{\bf x}-{\bf y}|$.

This general statement, in particular case that will be relevant for us, can be presented as follows \cite{patashinskii,cardy}.
Let  $\Phi$ (called scaling field)  be some product of massless free fields.
 Then
\begin{eqnarray}
\label{rcl}
&&  :\Phi_i({\bf x}):\times:\Phi_j({\bf y}):
  =  :\Phi_i({\bf x}) \Phi_j({\bf x}): +\nonumber \\
&& +  \sum_{1-contraction\ [({\bf x},i)({\bf y},j)]} \Delta_{\cal{AB}}(|{\bf x}-{\bf y}|)  :\Phi_i({\bf x})\Phi_j({\bf x}): \nonumber \\
&& +\sum_{2-contractions\ [({\bf x},i)({\bf y},j)],[({\bf x},i')({\bf y},j')]}\
\Delta_{\cal{AB}} (|{\bf x}-{\bf y}|)\nonumber\\
 &&  \Delta_{\cal{A}'\cal{B}'} (|{\bf x}-{\bf y}|):\Phi_i({\bf x})\Phi_j({\bf x}):+ \dots,
\end{eqnarray}
where
$:X :$  stands for normal ordered operator  $X$, and
\begin{eqnarray}
\Delta_{\cal{AB}}(x)=\frac{\delta_{AB}}{4\pi}\frac{\Gamma(\sigma)}{\pi^{\sigma}}\frac{1}{x^{2\sigma}},\;\;\;\sigma=d/2-1
\end{eqnarray}
 is the propagator of the free fields.(Furthe on, not to clutter notation, we'll omit the colon signs, where it can not lead to confusion.)

Let us consider a fixed point Hamiltonian $H^*$ which is perturbed by a number of scaling fields, so that the partition function is \cite{cardy}
\begin{eqnarray}
Z=\text{Tr}\; \exp\left\{-\int d^d{\bf r}\left[H^*+\sum_ia_c^{x_i}G_i\Phi_i({\bf r})\right]\right\},
\end{eqnarray}
where $x_i$ is the appropriate natural scaling dimension, and   microscopic cut-off   $a_c$  is  implied in the integral.
Expanding in the powers of coupling we obtain
\begin{eqnarray}
&&Z=Z^*\left[1-\sum_i a_c^{x_i-d}G_i\int d^d{\bf r}\langle\Phi_i({\bf r})\rangle\right.\\
&&\left.+\frac{1}{2}\sum_{ij}a_c^{x_i+x_j-2d}G_iG_j\int d^d{\bf r}_1d^d{\bf r}_2\langle\Phi_i({\bf r}_1)\Phi_j({\bf r}_2)\rangle-\dots\right],\nonumber
\end{eqnarray}
where all correlation functions are to be evaluated with respect to the fixed point Hamiltonian $H^*$.

We implement the RG by changing the microscopic cut-off from $a_c$ to $(1+d \ell)a_c$ and asking how the couplings $G_i$
should be changed to preserve the partition function $Z$. The answer is given by the perturbative RG equations  \cite{cardy}
\begin{eqnarray}
\label{sc}
\frac{dG_k}{d\ell}=(d-x_k)G_k
-\sum_{ij}c_{kij}G_iG_j+\dots,
\end{eqnarray}
where summation is with respect to all pairs $i,j$ such, that $\Phi_k$ appears in the product $\Phi_i\Phi_j$, as the result of contraction(s) (there shold be at least one).
The coefficients $c_{kij}$ depend upon the specific realization of renormalization procedure and   typically are presented containing multipliers expressed  through   the area of the hypersphere of unit radius in $d$ dimensions, $\pi$ and things like this, which appear as the result of calculation of loop integrals. However,  these multipliers are the same for all coefficients with the same number of contractions (loops). Because only ratio of the coefficients  is important,  if all the terms in Eq. (\ref{sc}) contain the same number of contractions (as will be in our case), we can always make all relevant $c_{kij}$ equal to 1 by appropriate rescaling of $G_i$. \cite{cardy}.

\section{Quenched scalar disorder and cubic anisotropy}
\label{quenched}

\subsection{Replica method}

Consider the $d$-dimensional   system described by the $n$-component order parameter $\phi_i({\bf r})$ $(i=1,2\dots,n)$ with  the symmetry  explicitly broken from $O(n)$ to cubic and quenched random scalar disorder. Combining the well known results we may describe the system in vicinity of
the critical points  by the following continuous Hamiltonian  \cite{ma,cardy,dotsenko}:
\begin{eqnarray}
\label{hamiltonian}
&&H[\delta\tau,\phi]=\int d^D{\bf r}\left\{\frac{1}{2}\sum_{a=1}^n\left(\nabla\phi_a({\bf r})\right)^2\right.\nonumber\\
&&+a_c^{-2}(\tau-\delta\tau({\bf r}))\sum_{a=1}^n\phi_a({\bf r})^2\\
&&\left.+ua_c^{d-4}\sum_{a,b=1}^n\phi_a({\bf r})^2\phi_b({\bf r})^2
+va_c^{d-4}\sum_{a=n}^n\phi_a({\bf r})^4\right\},\nonumber
\end{eqnarray}
where $a_c$ is a microscopic cut-off, $\tau =(T-T_c)/T_c$.
 (It is known that fluctuations of the other two coefficients in the Landau-Ginsburg functional do not influence critical behavior for small $\epsilon$. \cite{ma})
According to the replica method  one has to calculate the following partition function (fluctuations of the effective transition temperature $\delta\tau({\bf r})$ we assume  to be  Gaussian)
\begin{eqnarray}
&&Z_p=\overline{\left(\int \prod_{a=1}^nD\phi_a\exp\{-H[\delta\tau,\phi]\}\right)^p}\nonumber\\
&&=\int D\delta\tau({\bf r})\int \prod_{a=1}^n\prod_{\alpha=1}^pD\phi_a^{\alpha}\\
&&\exp\left\{-\frac{1}{4\Delta}\int d^d{\bf r} \left[a_c^{-d}(\delta\tau)^2-H[\delta\tau,\phi]\right]\right\},\nonumber
\end{eqnarray}
where the superscript $\alpha$ labels the replicas.

As it is well known, the scheme of the replica method can be described in the following steps \cite{dotsenko}. First, the measurable quantities we are interested in should be calculated for integer $p$. Second, the analytic continuation of the obtained functions of the parameter $p$ should be made for an arbitrary non-integer $p$. Finally, the limit $p\to 0$ should be taken.

After Gaussian integration over $\delta\tau({\bf r})$ one gets:
\begin{eqnarray}
Z_p=\int \prod_{\cal{A}} D\phi^{\cal{A}}\exp^{-\int d^d{\bf r}\left\{H_0[\phi]+H_{int}[\phi]\right\}},
\end{eqnarray}
where
\begin{eqnarray}
\label{z}
H_{int}[\phi]=a_c^{-2}\tau\sum_{\cal{A}}\Phi^{\cal {A}}+a_c^{d-4}\sum_{\cal{AB}} g_{\cal{AB}}\Phi^{\cal{AB}},
\end{eqnarray}
and
\begin{eqnarray}
\label{ham2}
g_{\cal{AB}}=(u+v\delta_{ab})\delta_{\alpha\beta}-\Delta;
\end{eqnarray}
calligraphic capital  letter  stands for a pair of  replica index  and vector index.

It will be convenient for us
 to  rewrite  Hamiltonian (\ref{z}) as
\begin{eqnarray}
\label{hamt94}
H_{int}=a_c^{-2}\tau \Phi^1 +a_c^{d-4}\left[u\overline{\Phi}-\Delta \Phi+v\widehat{\Phi}\right],
\end{eqnarray}
 where
\begin{eqnarray}
\label{phi1}
\Phi^1&=&\sum_{a\alpha}\left(\phi^{\alpha}_a\right)^2\\
\label{Phi}
\Phi&=&\sum_{ab\alpha \beta}\left(\phi^{\alpha}_a\right)^2\left(\phi^{\beta}_b\right)^2\\
\label{bar}
\overline{\Phi}&=&\sum_{ab\alpha}\left(\phi^{\alpha}_a\right)^2\left(\phi^{\alpha}_b\right)^2\\
\label{hat}
\widehat{\Phi}&=&\sum_{a\alpha}\left(\phi^{\alpha}_a\right)^4.
\end{eqnarray}

\subsection{Multiplication table}

To derive the RG equations we will need expansion (which  in the approximation used are probably better to call merging)  coefficients) for two types of products. These expansions  can be graphically presented as follows:
\begin{eqnarray}
\label{decouplinga}
&&\left(\phi_{\cal K}\right)^2\times\left(\phi_{\cal A}\right)^2\left(\phi_{\cal B}\right)^2
=\contraction{(}{\phi }{\phi)(xxx}{\phi}
\contraction[2ex]{(\phi x}{\phi}{)(\phi\phi)(xxxx}{\phi}
(\phi_{\cal K}\phi_{\cal K})(\phi_{\cal A}\phi_{\cal A})(\phi_{\cal B}\phi_{\cal B})\nonumber\\
&&+\contraction{(}{\phi}{\phi)(xxx}{\phi}
\contraction[2ex]{:(\phi }{\phi}{)(\phi xx}{\phi}
(\phi_{\cal K}\phi_{\cal K})(\phi_{\cal A}\phi_{\cal A})(\phi_{\cal B}\phi_{\cal B})+\text{permutations}
\end{eqnarray}
and
\begin{eqnarray}
\label{decouplingb}
&&\left(\phi_{\cal A}\right)^2\left(\phi_{\cal B}\right)^2\times \left(\phi_{\cal C}\right)^2\left(\phi_{\cal D}\right)^2\\
&&=\contraction{(}{\phi}{\phi)(\phi\phi)(xxxxx}{\phi}
\contraction[2ex]{(\phi\phi)(xxx}{\phi}{)(\phi\phi)(\phi xxx}{\phi}
(\phi_{\cal A}\phi_{\cal A})(\phi_{\cal B}\phi_{\cal B})(\phi_{\cal C}\phi_{\cal C})(\phi_{\cal D}\phi_{\cal D})\nonumber\\
&&+\contraction{(}{\phi}{\phi)(\phi\phi)(xxxxx}{\phi}
\contraction[2ex]{(\phi x}{\phi}{)(\phi\phi)(\phi\phi)(xxxxxx}{\phi}
(\phi_{\cal A}\phi_{\cal A})(\phi_{\cal B}\phi_{\cal B})(\phi_{\cal C}\phi_{\cal C})(\phi_{\cal D}\phi_{\cal D})\nonumber\\
&&+\contraction{(}{\phi}{\phi)(\phi\phi)(xxxxx}{\phi}
\contraction[2ex]{(\phi x}{\phi}{)(\phi\phi)(\phi xxxxx}{\phi}
(\phi_{\cal A}\phi_{\cal A})(\phi_{\cal B}\phi_{\cal B})(\phi_{\cal C}\phi_{\cal C})(\phi_{\cal D}\phi_{\cal D})
+\text{permutations},\nonumber
\end{eqnarray}
where we have  ignored the irrelevant terms \cite{cardy} and the field independent one.
Permutations means the diagrams which can be obtained from the drawn ones by interchanging field operators within the brackets and/or  ${\cal A}$ and ${\cal B}$ and/or ${\cal C}$ and ${\cal D}$ and/or (in the case of in Eq. (\ref{decouplingb}))
 ${\cal AB}$ and ${\cal CD}$.

Eqs. (\ref{decouplinga}), (\ref{decouplingb})  allow us to obtain the multiplication table for operators (\ref{phi1}) - (\ref{hat})
\begin{eqnarray}
\label{first}
\Phi^1\times\Phi &=& 4(2+pn)\Phi^1    \\
\Phi^1\times\overline{\Phi} &=&4(2+n)\Phi^1\\
\Phi^1\times\widehat{\Phi}&=& 12\Phi^1\\
\label{phii}
\Phi\times \Phi&=&8(8+pn)\Phi\\
\Phi\times\overline{\Phi} &=&8(2+n)\Phi+48\overline{\Phi}\\
\Phi\times\widehat{\Phi}&=& 24\Phi+48\widehat{\Phi}\\
\overline{\Phi}\times\overline{\Phi} &=&8(8+n)\overline{\Phi}\\
\overline{\Phi}\times\widehat{\Phi}&=&24\overline{\Phi}+48\widehat{\Phi} \\
\label{last}
\widehat{\Phi}\times\widehat{\Phi}&=&72\widehat{\Phi}.
\end{eqnarray}
(We  have omitted the propagators to draw attention just to the combinatoric multipliers.)

\subsection{RG equations}

For dimension $d$  slightly smaller than four, we can look
 for fixed points of the Hamiltonian (\ref{z}) in the vicinity
of the Gaussian one, that is consider  $H_{int}$ as a perturbation.
The perturbation theory  is actually expansion   with respect to parameter $\epsilon=4-d$. In the lowest approximation with respect to this parameter we should restrict ourselves by two first  terms in the r.h.s. of Eq. (\ref{sc}).

Thus taking into account our multiplication table (\ref{first}) - (\ref{last}) and
substituting the expansion coefficients  into Eq. (\ref{sc}) we obtain the RG equations
\begin{eqnarray}
\label{c1}
\frac{du}{d\ell}&=&\epsilon u -8\left[(8+n)u^2-12u\Delta+6u v\right]\\
\label{c2}
\frac{d\Delta}{d\ell}&=& \epsilon \Delta-8\left[(4+2n)u\Delta -(8+pn)\Delta^2+6\Delta v\right] \nonumber \\ \\
\label{c3}
\frac{dv}{d\ell}&=& \epsilon v-8[12u v-12 \Delta v+9v^2].
\end{eqnarray}
and
\begin{eqnarray}
\label{ttau5}
\frac{d\tau}{d\ell}=2\tau-8\left[(2+n) u-(2+pn)\Delta+3v\right]\tau.
\end{eqnarray}

The  RG equations coincide with those   of Ref. \onlinecite{aharony2} (for generalization including $\epsilon^2$ terms  see Refs. \onlinecite{lawrie} and  \onlinecite{sarkar}).

\section{Random direction of the anisotropy axis}
\label{simple}

The random-axis model
\begin{eqnarray}
\label{anisotropy}
H=-J\sum_{<ij>}J_{ij}\vec{S}_i\vec{S}_j-D_0\sum_i(\vec{x}_i\vec{S}_i)^2
\end{eqnarray}
was introduced by Harris et al. \cite{harris}
 to describe the magnetic
properties of amorphous alloys. In Eq. (\ref{anisotropy}) $\vec{S}_i$ is an $n$-component spin vector located at the lattice site $i$, $<ij>$ denotes a pair of spin sites, $J_{ij}$ is the exchange interaction,  $\hat{x}_i$ is a unit vector which points in the local
(random) direction of the uniaxial anisotropy at the
site $i$, and $_0$ is the anisotropy constant.

The Hamiltonian of the model in the continuum approximation and after the replica trick can be presented as \cite{aharony2}
\begin{eqnarray}
\label{hamt9}
&&H_{int}=a_c^{-2}\tau \sum_{\cal{A}}\Phi^{\cal{A}}
+a_c^{d-4}\left[\sum_{\cal{AB}} g_{\cal{AB}}\Phi^{\cal{AB}}+w\Psi\right],\nonumber\\
\end{eqnarray}
 where
\begin{eqnarray}
\label{ham297}
g_{\cal{AB}}=g_{ab}=u\delta_{\alpha\beta}-\Delta,
 \end{eqnarray}
and
\begin{eqnarray}
\label{b3}
\Psi=\sum_{ ab\alpha\beta}\phi^{\alpha}_a\phi^{\alpha}_b\phi^{\beta}_a\phi^{\beta}_b.
\end{eqnarray}
The connection between the parameters of the Hamiltonians (\ref{anisotropy}) and (\ref{hamt9}) will be of no interest to us.

It is
convenient to  rewrite  Hamiltonian (\ref{hamt9}) as
\begin{eqnarray}
\label{hamt97}
H_{int}=a_c^{-2}\tau \Phi^1 +a_c^{d-4}\left[\overline{\Phi}-\Delta \Phi+w\Psi\right],
\end{eqnarray}
 where $\Phi^1$, $\overline{\Phi}$, $\Phi$ and $\Psi$
 were defined above.

Using Eq. (\ref{decouplingb}) we obtain additional lines of the
multiplication table necessary for obtaining the RG equations in the case considered.
\begin{eqnarray}
\label{ph}
&&\Phi^1\times \Psi=4(1+p+n)\Phi^1\\
&&\Phi\times \Psi=8(1+p+n)\Phi+48\Psi\\
\label{phipsi}
&&\overline{\Phi}\times \Psi=8\Phi+8(5+n)\overline{\Phi}+16\Psi\\
\label{phipsi6}
&&\Psi\times \Psi=8(4+p+n)\Psi+24\Phi.
\end{eqnarray}

Thus  we obtain the RG equations
\begin{eqnarray}
\label{c1b}
&&\frac{du}{d\ell}=\epsilon u -8\left[(8+n)u^2-12u\Delta +2(5+n)uw\right]\\
\label{c2b}
&&\frac{d\Delta}{d\ell}= \epsilon \Delta-8\left[(4+2n)u\Delta -(8+pn)\Delta^2 \right.\nonumber\\
&&\left. -2uw+2(1+p+n)\Delta w-3w^2\right]\\
\label{c3b}
&&\frac{dw}{d\ell}= \epsilon w -8\left[4uw-12\Delta w+(4+p+n)w^2\right].
\end{eqnarray}
The analog of Eqs. (\ref{ttau5})  is
\begin{eqnarray}
\label{ttau2}
\frac{d\tau}{d\ell}=2\tau-8\left[(2+n) u-(2+pn)\Delta+(1+p+n)w\right]\tau.\nonumber\\
\end{eqnarray}
Eqs. (\ref{c1b}) - (\ref{ttau2})  exactly coincide with those from Ref. \cite{aharony2}.

Again, finally we should go  in Eqs. (\ref{c1b})-(\ref{ttau2}) to the limit $p\to 0$.

\section{Replica symmetry breaking}
\label{rsb}

Consider first the case of scalar disorder from  Section \ref{quenched}. In this case there is one replica non-diagonal scaling field -- $\Phi$.
Hence possible RSB will be taken into account if we  generalize the  Hamiltonian (\ref{hamt94}) in the following way.
\begin{eqnarray}
\label{hamt94b}
H_{int}=a_c^{-2}\tau \Phi^1 +a_c^{d-4}\left[u\overline{\Phi}-\sum_{\alpha\beta}\Delta_{\alpha\beta} \Phi_{\alpha\beta}+v\widehat{\Phi}\right],
\end{eqnarray}
 where
\begin{eqnarray}
\label{Phir}
\Phi_{\alpha\beta}=\sum_{ab}\left(\phi^{\alpha}_a\right)^2\left(\phi^{\beta}_b\right)^2.
\end{eqnarray}

We have to make more specific the lines containing $\Phi$ in our multiplication table (\ref{first}) - (\ref{last}):
\begin{eqnarray}
\label{first2}
&&\Phi^1\times\sum_{\alpha\beta}\Delta_{\alpha\beta}\Phi_{\alpha\beta} = 4\sum_{a\alpha}\left(2\Delta_{\alpha\alpha}+n\sum_{\beta}\Delta_{\alpha\beta}\right)\left(\phi_a^{\alpha}\right)^2 \nonumber \\  \\
\label{phii2}
&&\sum_{\alpha\beta}\Delta_{\alpha\beta}\Phi_{\alpha\beta}\times\sum_{\gamma\delta}\Delta_{\gamma\delta} \Phi_{\gamma\delta}=8\sum_{\alpha\beta}\Bigg[4\Delta_{\alpha\beta}^2\nonumber\\
&&+2\left(\Delta_{\alpha\alpha}+\Delta_{\beta\beta}\right)\Delta_{\alpha\beta}+n\sum_{\gamma}\Delta_{\alpha\gamma}\Delta_{\gamma\beta}\Bigg]\Phi_{\alpha\beta}\\
&&\sum_{\alpha\beta}\Delta_{\alpha\beta}\Phi_{\alpha\beta}\times\overline{\Phi} =8\sum_{\alpha\beta}\left(2+n\right)\Delta_{\alpha\beta}\Phi_{\alpha\beta}\nonumber\\
&&+48\sum_{ab\alpha}\Delta_{\alpha\alpha}\left(\Phi^{\alpha}_a\right)^2\left(\Phi^{\alpha}_b\right)^2\\
&&\sum_{\alpha\beta}\Delta_{\alpha\beta}\Phi_{\alpha\beta}\times\widehat{\Phi}= 24\sum_{\alpha\beta}\Delta_{\alpha\beta}\Phi_{\alpha\beta}
+48\sum_{a\alpha}\Delta_{\alpha\alpha}\left(\Phi^{\alpha}_a\right)^4.\nonumber\\
\end{eqnarray}
Substituting the results from our expandeed multiplication table int Eq	. (\ref{sc}) we obtain the RG equations. We shall study these equations assuming that the matrix $\Delta_{\alpha\beta}$ 
has a general Parisi RSB structure, and in the limit $p\to 0$ is parameterized in terms of its
diagonal elements $\widetilde{\Delta}$ and the off-diagonal function $\Delta(x)$ defined in the interval $0<x<0$ (which can be presented as $\Delta=(\tilde{\Delta},\Delta(x)$). \cite{dotsenko}.
Due to such parametrization  we immediately recover
Eqs. (\ref{c1}), (\ref{c3}) and  (\ref{ttau5}) (the last one with $p=0$), with $\Delta$ substituted by $\tilde{\Delta}$.

Where the standard technique of the Parisi RSB algebra is substantially different from ordibnary matrix algebra is  product of matrices  \cite{parisi}
(and we have such product in the r.h.s. of Eq. (\ref{phii2})). The  definition of the product of  Parisi matrices  is as follows.
Let $a=(\tilde{a},a(x))$, $b=(\tilde{b},b(x))$, $c=(\tilde{c},c(x))$, and $c=ab$. Then
\begin{eqnarray}
\label{hr}
&&\tilde{c}=\tilde{a}\tilde{b}-\int_0^1dxa(x)b(x)\\
\label{en}
&&c(x)=\left(\tilde{a}-\int_0^1dya(y)\right)b(x)+\left(\tilde{b}-\int_0^1dyb(y)\right)a(x)\nonumber\\
&&-\int_0^xdy[a(x)-a(y)][b(x)-b(y)].
\end{eqnarray}
Thus we recover Eq. (\ref{c2}) (with $p=0$), only this time $\Delta$ is not a number but a Parisi matrix.
We generalized RG equations from Ref. \onlinecite{dotsenko2} for the case of cubic
 anisotropy present in the model.

For the case of random anisotropy axis the situation is very much similar.  In this case there are two replica non-diagonal scaling fields $\Phi$ and $\Psi$. Hence the Hamiltonian which 
takes into account possible symmetry breaking should be written in the following form.
\begin{eqnarray}
\label{hamt997}
&&H_{int}=a_c^{-2}\tau \Phi^1 +a_c^{d-4}\left[u\overline{\Phi}-\sum_{\alpha\beta}\Delta_{\alpha\beta} \Phi_{\alpha\beta}\right.\nonumber\\
&&\left.+\sum_{ ab\alpha\beta}w_{\alpha\beta}\phi^{\alpha}_a\phi^{\alpha}_b\phi^{\beta}_a\phi^{\beta}_b\right],
\end{eqnarray}
where $\Delta$ and $w$ are both Parisi matrices.
Repeating the derivation from above we again recover  Eqs. (\ref{c1b}),  (\ref{ttau2})  (the last one with $p=0$), where  $\widetilde{\Delta}$ substitutes for $\Delta$, and . $\widetilde{w}$ substitutes for $w$,
and Eqs. (\ref{c2b}), (\ref{c3b})  (with $p=0$), where product of the matrices $\Delta$ and $w$ is understood according to Eqs. (\ref{hr}) and (\ref{en}). 

The analysis of the fixed points of the RG equations obtained in this Section is left
for  consideration in future.

\section{Appendix}

Phase transitions are described by the stable fixed points of the RG equations.
In this Appendix  for pedagogical purposes we  present analysis of the fixed points of the replica symmetric RG equations for the
quenched scalar disorder and their stability.

\noindent
Eqs. (\ref{c1}) - (\ref{c3}) have 7 fixed points \cite{aharony2}:

\noindent
(i) Gaussian fixed point: $(u^*,\Delta^*,v^*)=(0,0,0)$

\noindent
(ii) Pure Heisenberg fixed point: $(u^*,\Delta^*,v^*)=(\epsilon/8(8+n),0,0)$;

\noindent
(iii) Pure Ising fixed point: \cite{cardy}  $(u^*,\Delta^*,v^*)=(0,0,\epsilon/72)$;

\noindent
(iv) Pure cubic fixed point: \cite{cardy} $(u^*,\Delta^*,v^*)=(\epsilon/24n,0,(n-4)\epsilon/72n)$;

\noindent
(v) Random Heisenberg fixed point: $(u^*,\Delta^*,v^*)=(\epsilon/32(n-1),(4-n)\epsilon/128(n-1),0)$;

\noindent
(vi) Random cubic fixed point:  $(u^*,\Delta^*,v^*)=(1/48(n-2),(4-n)/192(n-2),(n-4)/144(n-2))$;

\noindent
(vii) Nonphysical fixed point:   $(u^*,\Delta^*,v^*)=(0,-\epsilon/64,0)$.

\noindent
The nonphysically of the last fixed point  is due to the fact that  the value of $\Delta^*$
is negative; on the other hand,  being a mean square value of
quenched disorder fluctuations, it is only positive defined \cite{dotsenko}.

Notice that the  random Heisenberg fixed point is physically meaningful for $4\geq n>1$, and
 the random cubic fixed point is physically meaningful for $4\geq n>2$.

To analyze stability of the fixed points we  assume
\begin{eqnarray}
u=u^*+\delta u,\;\;\; \Delta=\Delta^*+\delta \Delta,\;\;\; v=v^*+\delta v
\end{eqnarray}
and linearize the RG equations.

In
 the vicinity of the Gaussian fixed point we obtain
\begin{eqnarray}
\frac{d}{d\ell}\left(\begin{array}{c}\delta u \\ \delta \Delta \\ \delta v\end{array}\right)=
\epsilon \left(\begin{array}{ccc} 1 & 0 &0 \\ 0 &  1 & 0\\ 0 & 0 & 1\end{array}\right)
\left(\begin{array}{c}\delta u \\ \delta \Delta \\ \delta v\end{array}\right).
\end{eqnarray}
The Gaussian  fixed point  is stable for $d> 4$.

In the vicinity of the pure Heisenberg  fixed point we obtain
\begin{eqnarray}
\frac{d}{d\ell}\left(\begin{array}{c}\delta u \\ \delta \Delta \\ \delta v\end{array}\right)=
\frac{\epsilon}{8+n}\left(\begin{array}{ccc} -(8+n)  & 12 & -6 \\
0  &  4-n & 0\\
0 & 0 & n-4\end{array}\right)
\left(\begin{array}{c}\delta u \\ \delta \Delta \\ \delta v\end{array}\right).\nonumber\\
\end{eqnarray}
This  fixed point  is  unstable both for $n>4$ and $n<4$.

In the vicinity of the pure Ising fixed point we obtain
\begin{eqnarray}
\label{ising}
\frac{d}{d\ell}\left(\begin{array}{c}\delta u \\ \delta \Delta \\ \delta v\end{array}\right)
=\frac{\epsilon}{3}\left(\begin{array}{ccc} 1  & 0 & 0\\
0  &  1 & 0\\
-4 & 4 & -6 \end{array}\right)
\left(\begin{array}{c}\delta u \\ \delta \Delta \\ \delta v\end{array}\right).
\end{eqnarray}
This  fixed point  is always unstable.

In the vicinity of the pure cubic fixed point we obtain
\begin{eqnarray}
\label{three}
&&\frac{d}{d\ell}\left(\begin{array}{c}\delta u \\ \delta \Delta \\ \delta v\end{array}\right)\\
&&=\frac{\epsilon}{3n}\left(\begin{array}{ccc} -(8+n)  & 12 & -6\\
0  &  4-n & 0\\
4(4-n) & -4(4-n)  & 3(4-n) \end{array}\right)
\left(\begin{array}{c}\delta u \\ \delta \Delta \\ \delta v\end{array}\right).\nonumber
\end{eqnarray}
The three eigenvalues $\mu_1$, $\mu_2$ and $\mu_3$ of the system (\ref{three}) are
\begin{eqnarray}
\mu_1= -\epsilon,\;\;\;\mu_2=\frac{(4-n)\epsilon}{3n},\;\;\;\mu_3=4-n.
\end{eqnarray}
This  fixed point  is stable for $d<4$ and $n>4$.

In the vicinity of the random Heisenberg fixed point we obtain
\begin{eqnarray}
\label{trick}
&&\frac{d}{d\ell}\left(\begin{array}{c}\delta u \\ \delta \Delta \\ \delta v\end{array}\right)
=\frac{\epsilon}{4(n-1)}\\
&&\left(\begin{array}{ccc} -(8+n)  & 12 & -6/(8+n)\\
(1+n/2)(n-4)  &  2(4-n) & -6/(8+n)\\
0 & 0 & n-4 \end{array}\right)
\left(\begin{array}{c}\delta u \\ \delta \Delta \\ \delta v\end{array}\right).\nonumber
\end{eqnarray}
The three eigenvalues of the system (\ref{trick}) $\mu_1$, $\mu_2$ and $\mu_3$ are
\begin{eqnarray}
\mu_1=-\epsilon, \;\;\;\mu_2=\frac{(n-4)\epsilon}{4(n-1)}, \;\;\;\mu_3=n-4.
\end{eqnarray}
Hence the   fixed point  is stable for $d<4$ and $n <4$.

In the vicinity of the random cubic fixed point we obtain
\begin{eqnarray}
\label{general2}
&&\frac{d}{d\ell}\left(\begin{array}{c}\delta u \\ \delta \Delta \\ \delta v\end{array}\right)
=\left(\begin{array}{ccc}  -8(8+n)u^*  & 96u^* & -48u^*\\
-16(2+n)\Delta^*  &  64\Delta^*  & -48\Delta^*\\
-96v^* & 96v^* & -72v^* \end{array}\right) \nonumber\\
&&\cdot\left(\begin{array}{c}\delta u \\ \delta \Delta \\ \delta v\end{array}\right)
=\frac{\epsilon}{12(n-2)}\\
&&\left(\begin{array}{ccc}  -(8+n)  & 24 & -12\\
(2+n)(n-4)  &  4(4-n)  & 3(n-4)\\
8(4-n) & 8(n-4) & 6(4-n) \end{array}\right)
\left(\begin{array}{c}\delta u \\ \delta \Delta \\ \delta v\end{array}\right).\nonumber
\end{eqnarray}
From the fact that determinant of the matrix in Eq. (\ref{general2})  is positive
we come to the conclusion that at least one of the eigenvalues of the matrix  is positive, hence the random cubic fixed point is unstable.

\begin{figure}[h]
\includegraphics[width= .7\columnwidth, clip=true]{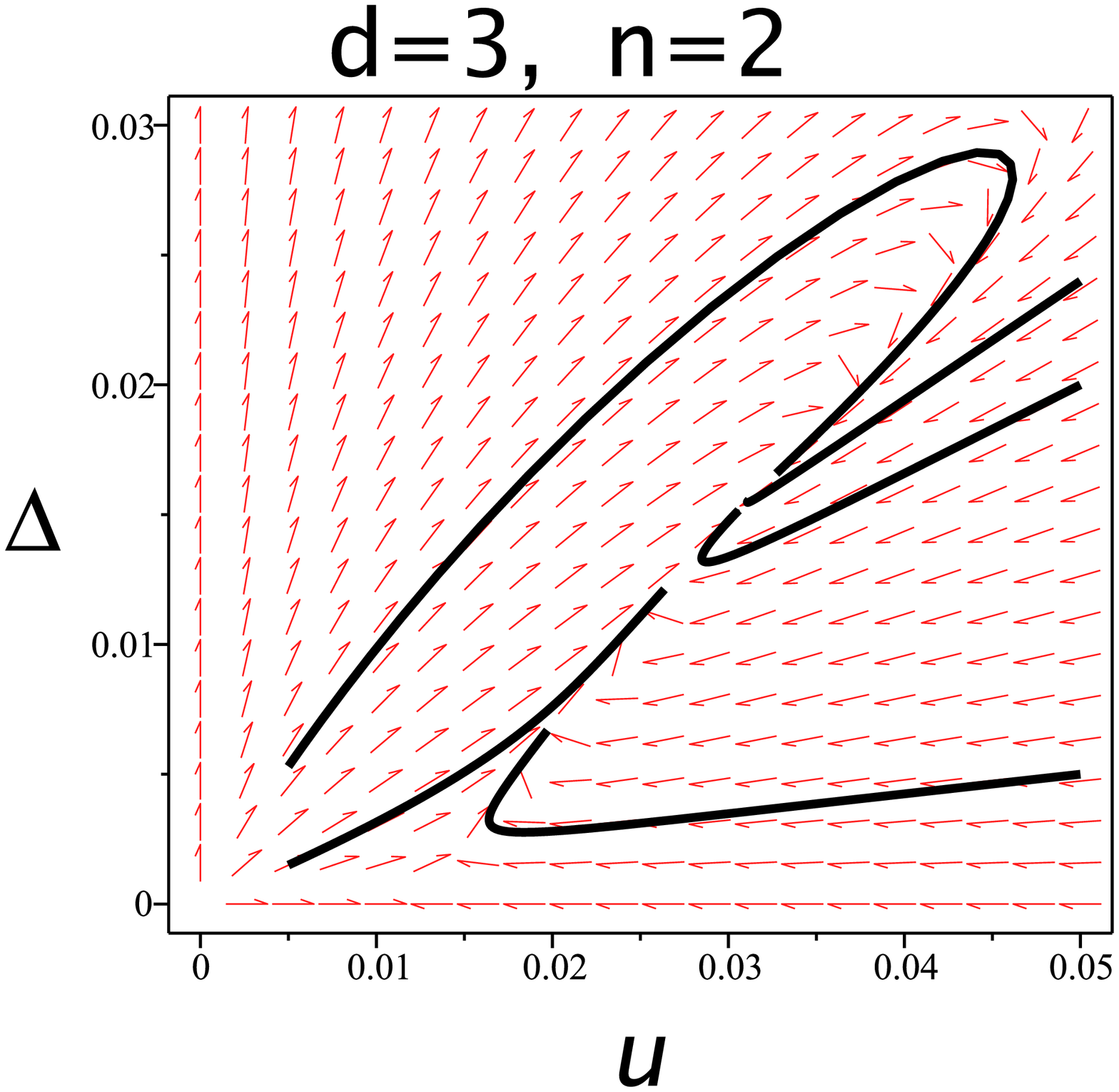}
\includegraphics[width= .7\columnwidth, clip=true]{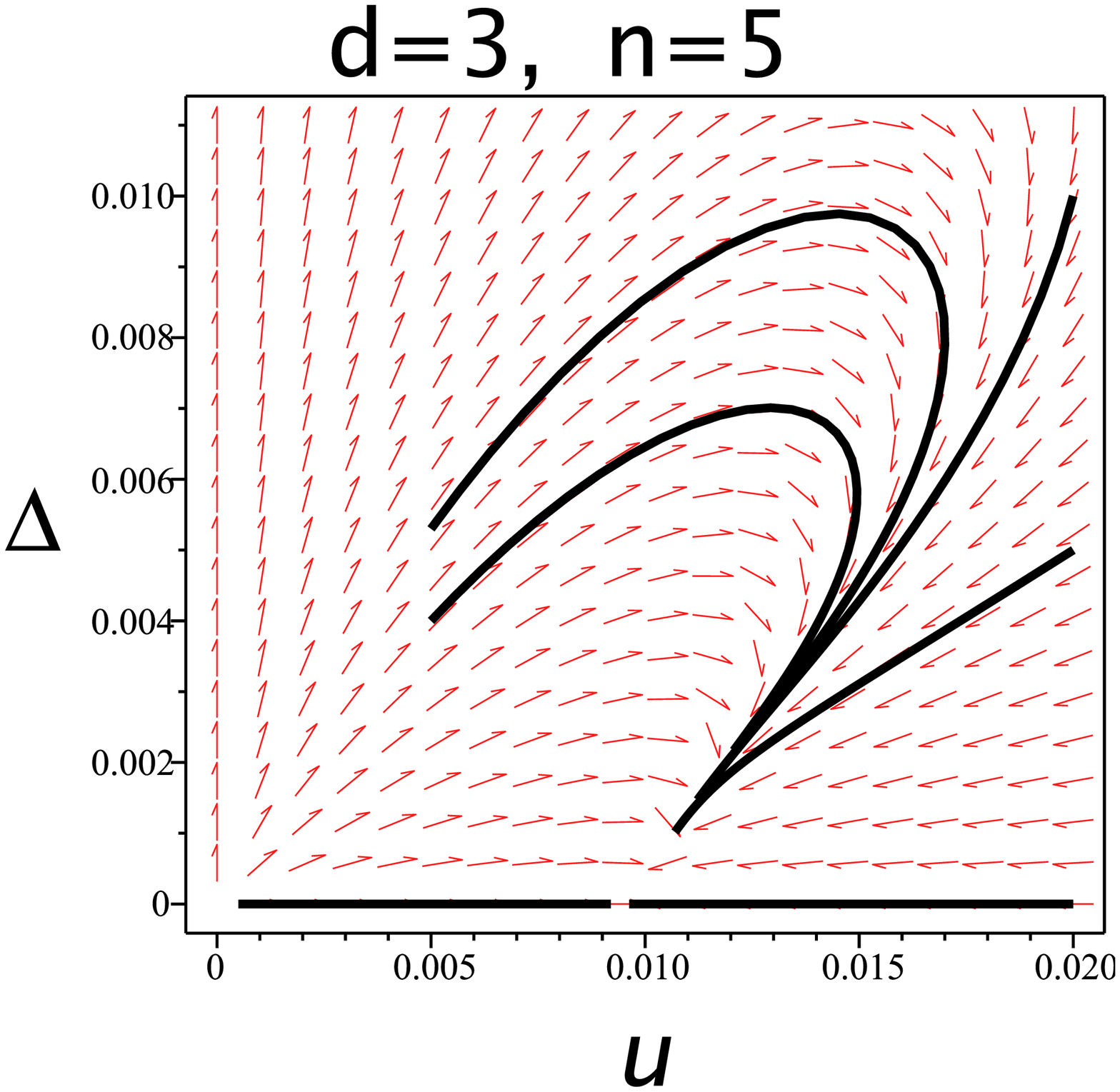}
\caption{\label{fig:d3n2} Phase portraits of the RG equations for $O(n)$ symmetric $\phi^4$ model in $d$ dimensions with quenched disorder (Eqs. (\ref{c1}),(\ref{c2}) with $p=0$ and $v=0$). }
\end{figure}

\begin{figure}[h]
\includegraphics[width= .7\columnwidth, clip=true]{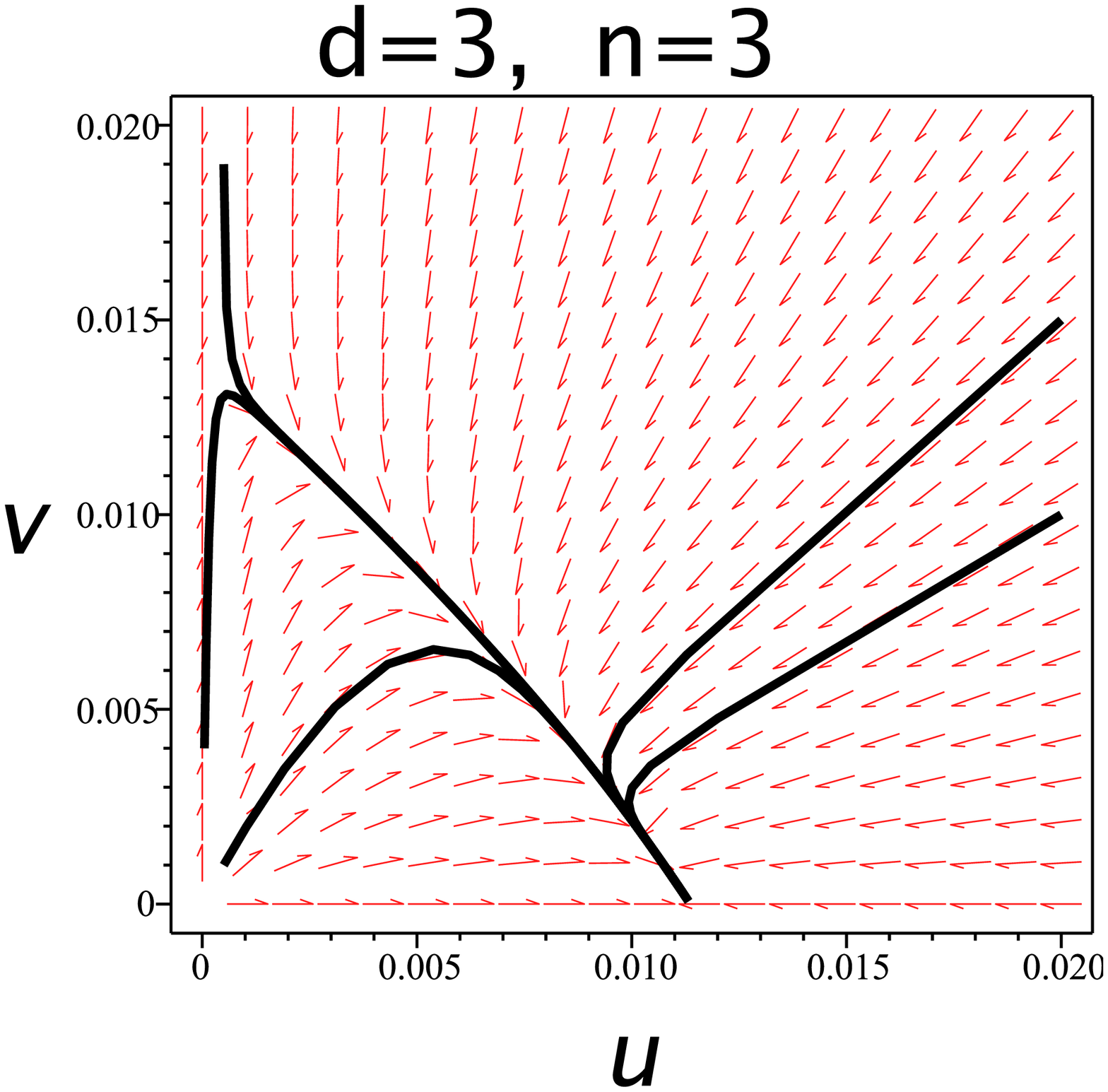}
\includegraphics[width= .7\columnwidth, clip=true]{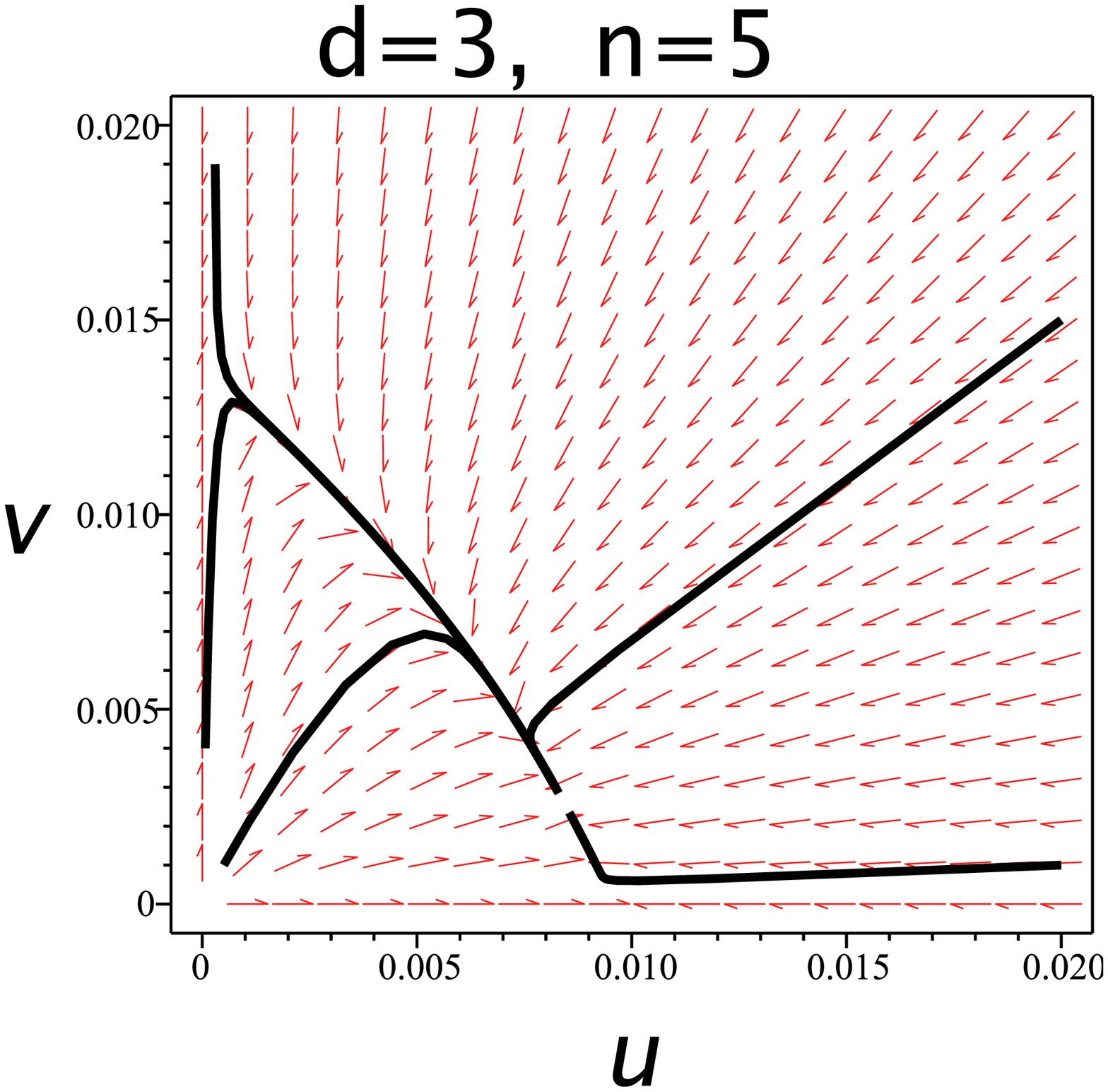}
\caption{\label{fig:cubicd3}  Phase portraits of the RG equations for pure $\phi^4$ model with cubic anisotropy in $d$ dimensions  (Eqs. (\ref{c1}),(\ref{c3})  with $p=0$ and $\Delta=0$). }
\end{figure}

Phase portraits of the system  (\ref{c1}),(\ref{c2}), (\ref{c3}) (for $p=0$) for $d=3$ are presented: without cubic anisotropy  ($v=0$) --  on   Fig. \ref{fig:d3n2},
and without disorder ($\Delta=0$)  -- on Fig. \ref{fig:cubicd3}. We see the stable  random Heisenberg and  pure cubic fixed points for  $n <4$ and   $n >4$ respectively, and the  unstable Gaussian and pure Ising fixed points.

The critical exponent $\nu$ is found from the RG equation for $\tau$: \cite{aharony2}
\begin{eqnarray}
\label{nunu}
\nu^{-1}=\lambda_{\tau}= \left.\frac{d\ln\tau}{d\ell}\right|_{(u,\Delta,v)=(u^*,\Delta^*,v^*)}\nonumber\\
=2-8\left[(2+n) u^*+3v^*-2\Delta^*\right].
\end{eqnarray}
For the random Heisenberg fixed point
$\nu_R=\frac{1}{2}+\frac{3n\epsilon}{32(n-1)}$,
and for the pure cubic fixed point
$\nu_{PC}=\frac{1}{2}+\frac{(n-1)\epsilon}{6n}$.
Since in linear to $\epsilon$ approximation there is no renormalization of the gradient term in Eq. (\ref{z}), for all fixed points
$\eta=0$.
Other critical exponents can  be obtain from the two calculated ones using scaling relations \cite{ma}.

\section{Acknowledgements}

We see our modest contribution as one more illustration to the famous saying of Leopold Kronecker:   `Die ganzen Zahlen hat der liebe Gott gemacht, alles andere ist Menschenwerk' (`God made the integers, all else is the work of man').

One of the authors (E.K.) cordially thanks  for the hospitality extended to him during
his stay: Max-Planck-Institut fur Physik komplexer Systeme, where the work was initiated, and
Center for Theoretical Physics of Complex Systems, where the work continued.

Discussions with A. Aharony, J. Cardy, J. Holland, I. D. Lawrie, F. Pollmann, N. Sarkar, and A. Sinner   are gratefully acknowledged.

\end{document}